\documentclass[%
 reprint,
 amsmath,amssymb,
 aps,
]{revtex4-1}

\usepackage{graphicx}
\usepackage{dcolumn}
\usepackage{bm}
\usepackage{float} 
\usepackage{multirow}

\usepackage{subfig}

\newcommand{\STAB}[1]{\begin{tabular}{@{}c@{}}#1\end{tabular}}

\newcommand*{\affaddr}[1]{#1} 
\newcommand*{\affmark}[1][*]{\textsuperscript{#1}}
\usepackage[USenglish]{babel}
\begin{document}

\preprint{APS/123-QED}

\title{Long-term LHC Discovery Reach for Compressed Higgsino-like Models using VBF Processes} 

\author{
Nathalia Cardona\affmark[2], Andr\'es Fl\'orez\affmark[2], Alfredo Gurrola\affmark[1], Will Johns\affmark[1], Paul Sheldon\affmark[1], Cheng Tao\affmark[1]\\
\affaddr{\affmark[1] Department of Physics and Astronomy, Vanderbilt University, Nashville, TN, 37235, USA}\\
\affaddr{\affmark[2] Physics Department, Universidad de los Andes, Bogot\'a, Colombia}\\
}

\date{\today}

\begin{abstract}

The identity of Dark Matter (DM) is one of the most active topics in particle physics today. Supersymmetry (SUSY) is an extension of the standard model (SM) that could describe the particle nature of DM in the form of the lightest neutralino in R-parity conserving models. We focus on SUSY models that solve the hierarchy problem with small fine tuning, and where the lightest SUSY particles ($\tilde{\chi}_{1}^{0}$, $\tilde{\chi}_{1}^{\pm}$, $\tilde{\chi}_{2}^{0}$) are a triplet of higgsino-like states, such that the mass difference $\Delta m(\tilde{\chi}^{0}_{2},\tilde{\chi}^{0}_{1})$ is 2–50 GeV. We perform a feasibility study to assess the long-term discovery potential for these compressed SUSY models with higgsino-like states, using vector boson fusion (VBF) processes in the context of proton-proton collisions at $\sqrt{s} = 13$ TeV, at the CERN Large Hadron Collider. Assuming an integrated luminosity of 3000 fb$^{-1}$, we find that stringent VBF requirements, combined with large missing momentum and one or two low-$p_{T}$ leptons, is effective at reducing the major SM backgrounds, leading to a 5$\sigma$ (3$\sigma$) discovery reach for $m(\tilde{\chi}^{0}_{2}) < 180$ $(260)$ GeV, and a projected 95\% confidence level exclusion region that covers $m(\tilde{\chi}^{0}_{2})$ up to 385 GeV, parameter space that is currently unconstrained by other experiments. 
\end{abstract}

\pacs{Valid PACS appear here}
\maketitle


\section{\label{sec:level1}Introduction}

The particle nature of Dark Matter (DM) \cite{Hinshaw:2012aka,Aghanim:2018eyx,DMevidence} remains an enigma in particle physics today. Supersymmetry (SUSY)~\cite{SUSY1,SUSY2,SUSY3,SUSY4,SUSY5,SUSY6} is a well motivated model that intends to address several open questions in the standard model (SM) of particle physics, including the particle nature of DM. SUSY restores the symmetry between fermionic matter fields and bosonic force carriers, by including superpartners of SM particles whose spins differ by one-half unit with respect to their SM partners. In R-parity conserving models, SUSY particles are pair-produced and their decay chains end with a stable and electrically neutral SUSY particle, commonly referred to as the lightest supersymmetric particle (LSP). In several SUSY models, the LSP and canonical DM candidate is the lightest neutralino ($\tilde{\chi}_{1}^{0}$)~\cite{GoldbergPaper}, which is a mixture of the wino, bino, and higgsino fields that form the SUSY partners of the SM $W$, $\gamma/Z$, and Higgs fields, respectively.

The ATLAS~\cite{Aad:2008zzm} and CMS~\cite{Chatrchyan:2008aa} experiments at the Large Hadron Collider (LHC) have an extensive physics program to search for SUSY. However, for all of its attractive features, there is as yet no direct evidence in support of SUSY. Stringent bounds have been placed on the colored SUSY sector, excluding gluino ($\widetilde{g}$), stop ($\widetilde{t}$), and sbottom ($\widetilde{b}$) masses up to 2.31 TeV, 1.25 TeV and 1.24 TeV, respectively, depending on the model~\cite{SusyColoredSearch,CMS13TeVSusyColoredAllHadSearch2016data,CMS13TeVSusyColoredAllHadSearch2016to2018data,CMS13TeVSusyMT2Search2016to2018data,CMS13TeVSusy1LMetJetsSearch2016to2018data, Aaboud:2018ujj, Aaboud:2019trc, Aaboud:ATLAS13TeVSusyColoredAllHadSearch2016to2018data, Aaboud:ATLAS13TeVAllHadStopSearch2016to2018data}. Searches for SUSY in the electroweak sector, considering Drell-Yan (DY) production mechanisms of order $\alpha_{EW}^{2}$ (electroweak coupling squared) followed by decays to one or more charged leptons and missing momentum, have also placed bounds on chargino ($\widetilde{\chi}^{\pm}_{1}$) and neutralino ($\widetilde{\chi}^{0}_{1/2}$) masses up to 650 GeV in certain models \cite{Sirunyan:2018ubx,Aaboud:2018jiw,Aaboud:2018sua,Sirunyan:2018iwl}.

The absence of SUSY signals in previous and current experiments, in particular at the LHC, point to the possibility that either the SUSY particles are too heavy to be probed with the current LHC energies, or that the SUSY particles are hidden in the phase space where sensitivity is limited due to experimental constraints.

In this paper we consider one such category of experimentally challenging scenarios, where the mass gaps between the LSP and the other charginos and neutralinos is small. In these {\it{compressed spectrum scenarios}}, the momenta available to the co-produced SM particles are small, resulting in ``soft'' decay products that are challenging to detect. 

Compressed mass spectra arise in several regions of the SUSY parameter space. As one example, in the stau-neutralino coannihilation region, the mass difference between the scalar superpartner of the $\tau$ lepton ($\widetilde{\tau}$) and the $\widetilde{\chi}^{0}_{1}$ must not exceed approximately 50 GeV in order to obtain a calculated relic DM abundance consistent with that measured by the WMAP and Planck Collaborations~\cite{Carena:2012gp, Gurrola2008Paper, ConnectingPPandCosmology}. To probe this region, non-standard production mechanisms and search techniques have been used, such as Vector Boson Fusion (VBF)~\cite{VBF1,DMmodels2,Khachatryan:2016mbu,VBFSlepton,VBFStop,VBFSbottom,Florez:2019tqr,VBF2,Sirunyan:2019zfq,VBFZprime,VBFHN, VBFSpin2, VBFAxion} or boosted topologies from the presence of associated initial state radiation jets (ISR)~\cite{isrstauPHYS,Sirunyan:2019mlu,Sirunyan:2017jix,Khachatryan:2014rra,Aaboud:2017phn}. 

The target of this paper are the so-called natural SUSY
models that solve the hierarchy problem with minimal fine tuning. In the minimal supersymmetric extension of the SM (MSSM), the masses of the bino, wino, and higgsino states are parameterized in terms of the SUSY breaking terms $M_{1}$ and $M_{2}$, and the superpotential higgsino mass parameter $\mu$, respectively. The phenomenology of the electroweakinos is largely driven by these three parameters, especially for large values of tan$\beta$. As pointed out in Refs.~\cite{Feng:2013pwa,HowieHiggsinoReachPaper}, ``naturalness'' imposes constraints on the masses of higgsinos and suggests that $|\mu|$ be near the weak scale while $M_{1}$ and/or $M_{2}$ be larger~\cite{SParticlesNaturalness, HiddenSUSYatTheLHC, NaturalSUSYEndures, NaturalSUSYwith125GeVHiggs}. In such a scenario, the lightest SUSY particles ($\tilde{\chi}_{1}^{0}$, $\tilde{\chi}_{1}^{\pm}$, $\tilde{\chi}_{2}^{0}$) are a triplet of higgsino-like states, where the mass difference $\Delta m$ between the states is small and effectively determined by the magnitude of $M_{1}$ or $M_{2}$ relative to $|\mu|$. The pure higgsino DM scenario with MeV scale $\Delta m$ is interesting from a cosmology perspective as it can be combined with a non-thermal Big Bang Cosmology model to give rise to a calculated WIMP relic abundance that is consistent with the WMAP and Planck measurements. However, it has been pointed out in several studies, for example in Refs.~\cite{IsNaturalSUSYExcluded}, that this scenario may be excluded by direct detection constraints from the PandaX-II~\cite{Wang:2020coa}, LUX~\cite{Akerib:2016vxi} and Xenon-1t~\cite{Aprile:2019jmx} experiments, and by bounds from Fermi-LAT/MAGIC observations of gamma rays from dwarf spheroidal galaxies. On the other hand, the neutral higgsinos could have larger mass splittings $\Delta m$ due to non-negligible gaugino mixing, and are not excluded by direct searches. This higgsino-like dark matter scenario can potentially be probed in high-energy proton-proton ($pp$) collisions at the LHC. Constraints on small $\Delta m$ SUSY scenarios were first established by the LEP experiments~\cite{Heister:2001nk,Abdallah:2003xe,Achard:2003ge,Abbiendi:2003ji}, where the lower bound on direct $\tilde{\chi}_{1}^{\pm}$ or $\tilde{\chi}_{2}^{0}$ production corresponds to $m(\tilde{\chi}_{1}^{\pm}) \approx 75$-$92.4$ GeV (103.5 GeV) for $\Delta m(\tilde{\chi}_{1}^{\pm},\tilde{\chi}_{1}^{0}) < 3$ GeV ($> 3$ GeV) and higgsino-like cross sections.

\begin{figure}
 \begin{center} 
 \includegraphics[width=0.4\textwidth]{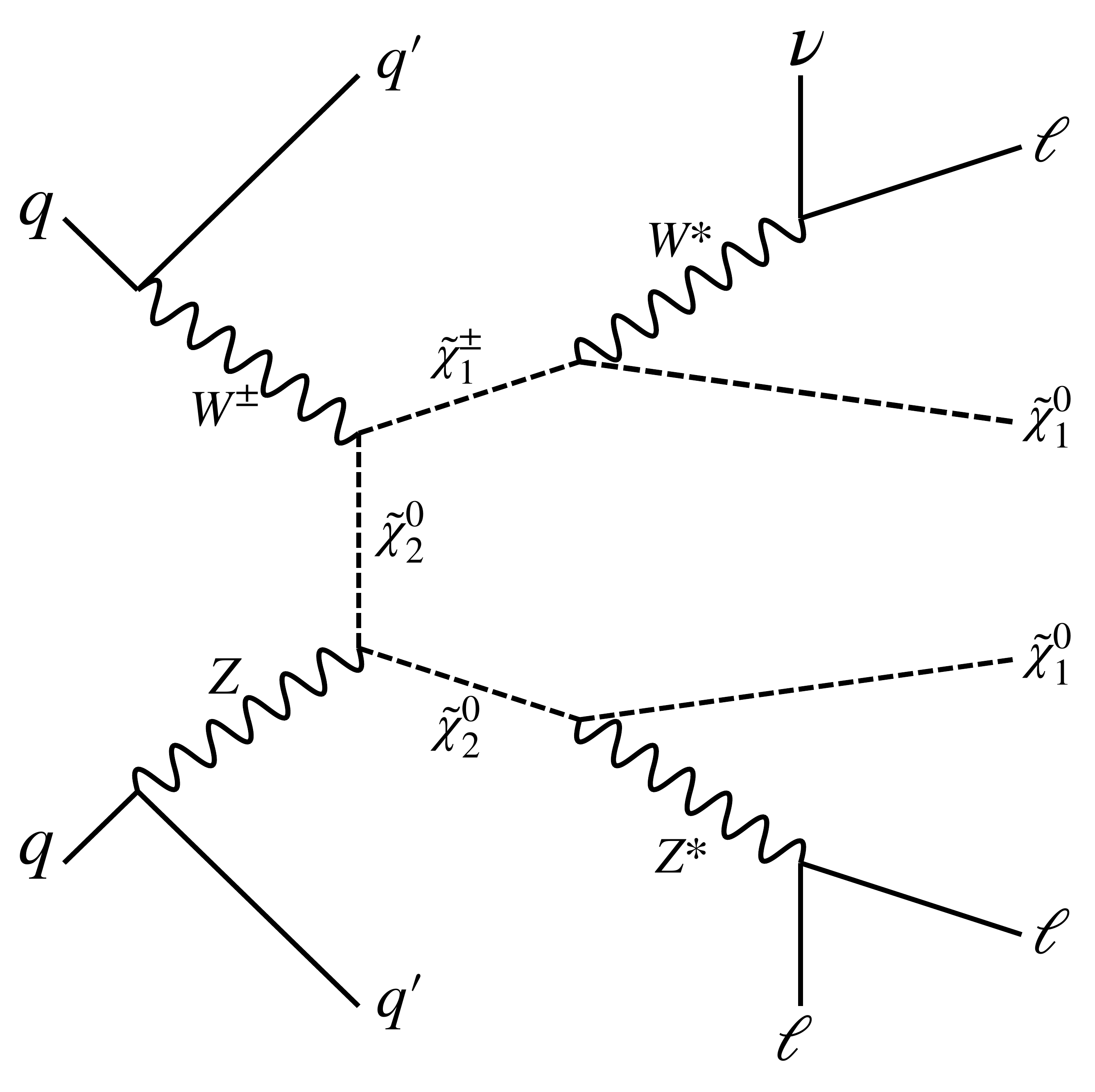}
 \end{center}
  \vspace{-0.7cm}
 \caption{Representative Feynman diagram of chargino-neutralino pair production through the t-channel $WZ$ fusion VBF process, followed by their decays to leptons and the LSP via virtual SM bosons.}
 \label{fig:feynDY}
 \end{figure}

\begin{figure}
 \begin{center} 
 \includegraphics[width=0.35\textwidth]{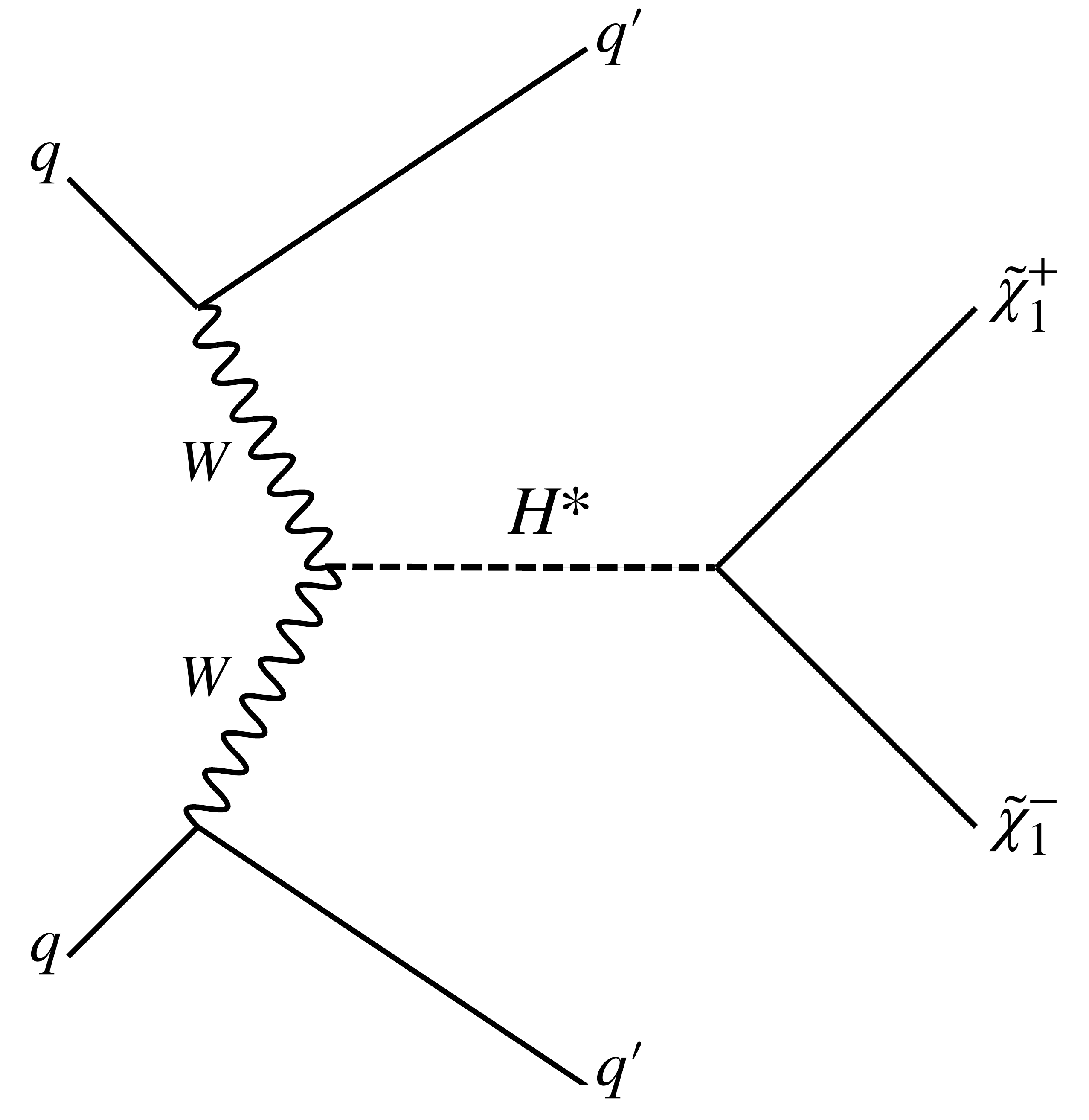}
 \end{center}
 \vspace{-0.5cm}
 \caption{Representative Feynman diagram depicting Higgsino production via the s-channel $WW$ fusion VBF process.}
 \label{fig:feynVBF}
 \end{figure}

The focus of this paper is a feasibility study to assess the long-term discovery potential for compressed SUSY models with higgsino-like states, using VBF processes in the context of $pp$ collisions at the High-Luminosity LHC (HLHC). In VBF processes, electroweak SUSY particles are pair-produced in association with two distinctive energetic jets. Figures~\ref{fig:feynDY} and~\ref{fig:feynVBF} show representative Feynman diagrams for the pair production of higgsino-like electroweakinos through VBF, in particular t-channel $WZ$ and s-channel $WW$ fusion. The VBF topology has proved to be a powerful experimental tool for compressed SUSY searches at the LHC due to the remarkable control over SM backgrounds, while also creating a kinematically boosted topology to help facilitate the reconstruction and identification of the soft decay products characterizing compressed mass spectra. We note that the long-term higgsino reach for non-VBF searches has recently been studied by the authors in Ref.~\cite{HowieHiggsinoReachPaper}. Our intent in this paper is to consider reasonable experimental conditions and uncertainties, to show how VBF higgsino searches can experimentally probe this challenging part of the phase space, complementing and expanding the reach with respect to previous studies. 

\section{Samples and simulation}

To assess the HLHC discovery reach for higgsino-like scenarios, samples of simulated events were used to determine the predicted signal and background yields, as well as the relevant kinematic distributions. Signal and background samples were both generated with MadGraph5\_aMC (v2.6.3.2)~\cite{MADGRAPH} considering $pp$ beams colliding with a center-of-mass energy of $\sqrt{s}=13$ TeV. The NNPDF3.0 NLO~\cite{Ball:2014uwa} parton distribution function (PDF) was used in the event generation of all samples. The parton-level events were then interfaced with the PYTHIA (v8.2.05)~\cite{Sjostrand:2014zea} package to include the 
parton showering (PS) and hadronization processes, while DELPHES(v3.4.1)~\cite{deFavereau:2013fsa} was used to simulate detector effects using the CMS detector geometric configurations and parameters for performance of particle reconstruction/identification. At the MadGraph parton level, jet pairs in signal (background) events were required to be well separated in $\eta$-$\phi$ space by using $|\Delta\eta(j_{1},j_{2})| >3.8$ ($\Delta R(j_{1},j_{2}) = \sqrt{\Delta \phi(j_{1}, j_{2})^{2}+\Delta \eta(j_{1},j_{2})^{2}} > 0.4$), where $\eta$ is the pseudorapidity ($\eta = -ln [tan (\theta/2)]$) and $\phi$ is the azimuthal angle. 
Similarly, at parton level, jets were required to have a minimum transverse momentum ($p_{T}$) of 30 GeV and $|\eta| < 5.0$. The cross sections in this paper are obtained with the aforementioned parton-level selections. The MLM algorithm~\cite{MLM} was used for jet matching and jet merging. The xqcut and qcut variables of the MLM algorithm, related with the minimal distance between partons and the energy spread of the clustered jets, were set to 30 and 45 as result of an optimization process requiring the continuity of the differential jet rate as a function of jet multiplicity. 

\begin{figure}[]
    \centering
    \includegraphics[width=0.48\textwidth]{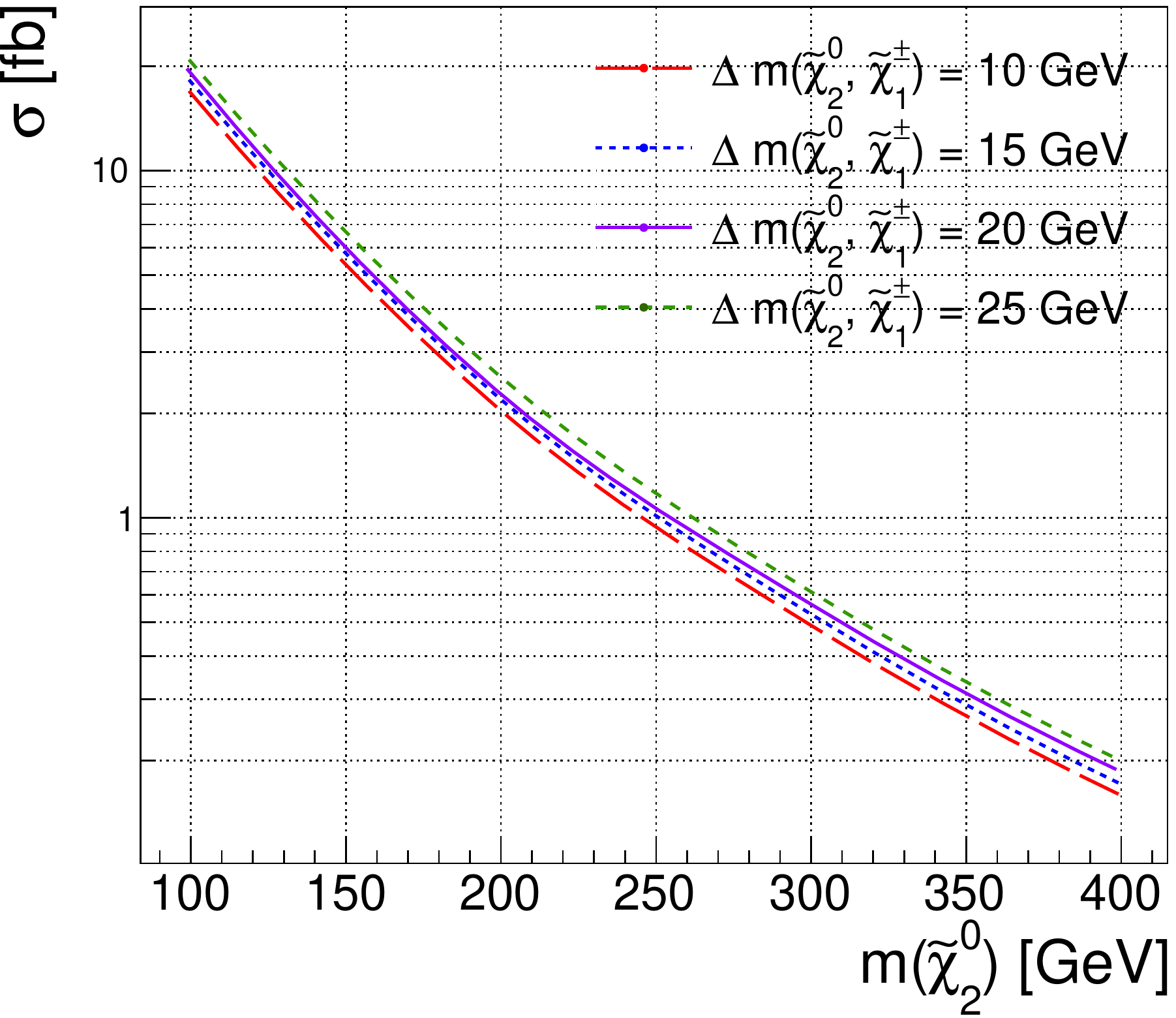}
    \caption{VBF chargino-neutralino pair-production cross section as a function of $m(\tilde{\chi}^{0}_{2})$.}
    \label{fig:cross_sec}
\end{figure}

The signal samples were produced in the context of the MSSM, considering R-parity conservation, and such that values of $M_{1}$, $M_{2}$, and $|\mu|$ result in a higgsino-like $\tilde{\chi}_{1}^{0}$. The signal scan was produced for small mass differences $\Delta m(\tilde{\chi}^{0}_{2},\tilde{\chi}^{0}_{1}) = 2$-$50$ GeV, to study compressed higgsino-like scenarios where experimental sensitivity at the LHC is currently limited. Prior higgsino searches from the CMS and ATLAS Collaborations have not exceeded the constraints established by the LEP experiments for $\Delta m(\tilde{\chi}^{0}_{2},\tilde{\chi}^{0}_{1}) < 3$ GeV, while the lower limit on $m(\tilde{\chi}^{0}_{2})$ is at 193 GeV for $\Delta m(\tilde{\chi}^{0}_{2},\tilde{\chi}^{0}_{1}) = 9.3$ GeV~\cite{ATLASCompressedSUSY13TeV2016to2018data,CMSCompressedSUSY13TeV2016to2018data}. It is noted that the MeV-scale mass gaps in the case of pure higgsino scenarios arise from radiative corrections. Therefore, the mass splittings considered in this paper were obtained by introducing mixing with wino or bino states via the gaugino mixing parameter $|\mu|$. The chargino mass was set to $m(\tilde{\chi}^{\pm}_{1}) = \frac{1}{2}m(\tilde{\chi}^{0}_{2}) + \frac{1}{2}m(\tilde{\chi}^{0}_{1})$ and all other SUSY particle masses were decoupled. It is noted that the value of $m(\tilde{\chi}^{0}_{2})$ in these higgsino-like scenarios can be slightly model dependent due to loop corrections. The assumption $m(\tilde{\chi}^{\pm}_{1}) = \frac{1}{2}m(\tilde{\chi}^{0}_{2}) + \frac{1}{2}m(\tilde{\chi}^{0}_{1})$ is accurate at leading order, but there may be small deviations due to higher order effects. This feature was pointed out by the authors in Ref.~\cite{HowieHiggsinoReachPaper}. However, this deviation from the assumed $m(\tilde{\chi}^{\pm}_{1}) = \frac{1}{2}m(\tilde{\chi}^{0}_{2}) + \frac{1}{2}m(\tilde{\chi}^{0}_{1})$ value is not very relevant for the VBF topology considered in this paper, a characteristic that was utilized to define the simplified model signal scans in Refs.~\cite{ATLASCompressedSUSY13TeV2016to2018data, CMSCompressedSUSY13TeV2016to2018data}. Signal events were simulated considering pure electroweak chargino-neutralino pair-production with two associated jets:  $pp\to \tilde{\chi}^{\pm}\tilde{\chi}_{1}^{0}jj$, $pp\to \tilde{\chi}^{\pm}\tilde{\chi}_{2}^{0}jj$, $pp\to \tilde{\chi}^{0}_{2}\tilde{\chi}_{1}^{0}jj$, $pp\to \tilde{\chi}^{\pm}_{1} \tilde{\chi}^{\pm}_{1}jj$ and $pp\to \tilde{\chi}^{0}_{2}\tilde{\chi}_{2}^{0}jj$. Figure~\ref{fig:cross_sec} shows the VBF signal production cross section as function of the $\tilde{\chi}^{0}_{2}$ mass, for different $\Delta m$ benchmark scenarios. Table~\ref{tab:cross_sec_examples} also lists the cross section values for some example benchmark points. For $m(\tilde{\chi}^{0}_{2})=100$ (200) GeV, the cross sections range from 11.6--20.8 (1.4--2.5) fb, depending on $\Delta m$. As the value of $\Delta m$ increases, the higgsino composition of $\tilde{\chi}_{2}^{0}$ and $\tilde{\chi}_{1}^{\pm}$ decreases, resulting in a larger cross section due to the increase in their wino composition. 

For the small $\Delta m$ higgsino-like scenarios considered in this paper, chargino-neutralino production is followed by $\tilde{\chi}^{0}_{2} \to l^{+}l^{-}\tilde{\chi}^{0}_{1}$ and $\tilde{\chi}^{\pm}_{1} \to l^{\pm}\nu_{l}\tilde{\chi}^{0}_{1}$ decays via virtual SM bosons, resulting in a final state of one or more charged leptons and missing transverse momentum ($p_{T}^{\textrm{miss}}$). We note that in the ``wino/bino scenario'' considered by the ATLAS and CMS Collaborations in Refs.~\cite{Sirunyan:2018ubx,Aaboud:2018jiw,Aaboud:2018sua,Sirunyan:2018iwl}, where $\tilde{\chi}_{2}^{0}/\tilde{\chi}_{1}^{\pm}$ are wino and $\tilde{\chi}_{1}^{0}$ is bino, VBF $\tilde{\chi}_{1}^{\pm}\tilde{\chi}_{2}^{0}$ production is the dominant process, composing about 60\% of the total signal cross section. On the other hand, Table~\ref{tab:winoVShiggsino} shows that the higgsino-like scenario considered in this paper contains a different composition of chargino-neutralino processes. This results in a different multiplicity of leptons in the final state, in comparison to the wino/bino scenario. 

The dominant sources of SM background are vector bosons ($W$,$Z$) produced in association with jets (referred to as $V$+jets), and pair production of top quarks ($t\bar{t}$). For the $V$+jets background, events with up to four jets were generated; additional jets may also result through the hadronization processes introduced by PYTHIA. The $V$+jets background samples are generated at next-to-leading-order (NLO) precision and include pure electroweak processes (such as vector boson fusion), taking into account interference effects between pure electroweak and mixed electroweak-QCD production. Subdominant processes were also considered, such as production of di-boson pairs in association with jets ($WW$+jets, $ZZ$+jets, and $WZ$+jets), single-top events, Higgs production, and tri-boson events. The single-top, Higgs, and tri-boson yields are grouped into the ``rare'' background category.

\begin{table}[]
\setlength{\tabcolsep}{7pt}
\label{tab:cross_sec_examples}
\caption{Signal cross sections for different benchmark points.}
\begin{tabular}{c|c|c}
\hline \hline 
$\Delta m(\tilde{\chi}^{0}_{2},\tilde{\chi}^{\pm}_{1})$ & $\{m(\tilde{\chi}^{0}_{2}), m(\tilde{\chi}^{\pm}_{1}),m(\tilde{\chi}^{0}_{1})\}$ & $\sigma$ [fb] \\
\hline
 \multirow{4}{*}{1} 
& \{100, 99, 98\}   & 11.57 \\
& \{200, 199, 198\} & 1.41 \\
& \{300, 299, 298\} & 0.34 \\
& \{400, 399, 398\} & 0.11 \\
\hline
\multirow{4}{*}{10} 
& \{100, 90, 80\}   & 16.50 \\
& \{200, 190, 180\} &2.01 \\
& \{300, 290, 280\} &0.48 \\
& \{400, 390, 380\} &0.16 \\
\hline
\multirow{4}{*}{15}
& \{100, 85, 70\}   & 17.75 \\
& \{200, 185, 170\} & 2.16 \\
& \{300, 285, 270\} & 0.52 \\
& \{400, 385, 370\} & 0.17 \\
\hline
\multirow{4}{*}{20}
& \{100, 80, 60\}   & 19.07 \\
& \{200, 180, 160\} & 2.32 \\
& \{300, 280, 260\} & 0.56 \\
& \{400, 380, 360\} & 0.18 \\
\hline
\multirow{4}{*}{25}
& \{100, 75, 50\}   & 20.80 \\
& \{200, 175, 150\} & 2.53 \\
& \{300, 275, 250\} & 0.61 \\
& \{400, 375, 350\} & 0.20 \\

\hline \hline 

\end{tabular}
\end{table}

\begin{table}[]
\setlength{\tabcolsep}{7pt}
\label{tab:winoVShiggsino}
\caption{Chargino-neutralino contributions to the total VBF cross section. These values are obtained for a benchmark scenario with $m(\tilde{\chi}_{2}^{0}) = 100$ GeV and $m(\tilde{\chi}_{1}^{0}) = 95$ GeV.}
\begin{tabular}{l|c|c}
\hline \hline
Process & Wino $\tilde{\chi}_{1}^{\pm}$/$\tilde{\chi}_{2}^{0}$, Bino $\tilde{\chi}_{1}^{0}$ & Higgsino $\tilde{\chi}_{1}^{\pm}$/$\tilde{\chi}_{2}^{0}$/$\tilde{\chi}_{1}^{0}$ \\ \hline

$\tilde{\chi}_{1}^{\pm}$$\tilde{\chi}_{2}^{0}$ & 58.9\% & 26.5\% \\

$\tilde{\chi}_{1}^{\pm}$$\tilde{\chi}_{1}^{0}$ & $\ll 0.1$
\% & 36.6\% \\ 

$\tilde{\chi}_{1}^{\pm}$$\tilde{\chi}_{1}^{\mp}$ & 33.9\% & 19.4\% \\ 

$\tilde{\chi}_{2}^{0}$$\tilde{\chi}_{2}^{0}$ & 7.2\% & 1.7\% \\ 

$\tilde{\chi}_{2}^{0}$$\tilde{\chi}_{1}^{0}$ & $\ll 0.1$
\% & 13.4\% \\ 

$\tilde{\chi}_{1}^{0}$$\tilde{\chi}_{1}^{0}$ & $\ll 0.1$
\% & 2.5\% \\
\hline \hline

\end{tabular}
\end{table}

\section{Event selection criteria}

Although the compressed higgsino-like scenarios result in final states with up to four leptons (in the case of $\tilde{\chi}_{2}^{0}\tilde{\chi}_{2}^{0}$ production), the decay products have an average transverse momentum of $p_{T} \sim \Delta m / 3$, which makes it difficult to experimentally reconstruct and identify all of them. Therefore, we select a final state topology with either one or two soft light leptons (electrons or muons). Final states with a hadronically decaying $\tau$ lepton ($\tau_{h}$) are not considered due 
to known experimental difficulties reconstructing low-$p_{T}$ genuine $\tau_{h}$ candidates, namely that they do not produce a narrow energy flow in the detector, making them difficult to distinguish from quark or gluon jets~\cite{Sirunyan:2018pgf, Sirunyan:2018vhk, ZptautauCMS, LQbbtautau3CMS}.

In addition to the presence of one or two soft light leptons, we require large $p_{T}^{\textrm{miss}}$ due to the presence of boosted $\tilde{\chi}_{1}^{0}$'s in the final state and require the presence of jets consistent with the characteristics of a VBF process. Stringent requirements are placed on the $p_{T}$ of leptons, $p_{T}^{\textrm{miss}}$, and on the kinematic properties of the VBF dijet system in order to suppress SM backgrounds. 

The exact cut values for all selections are determined through an optimization process aimed at maximizing discovery reach. For this purpose we take a simple approach to defining signal significance $z = \frac{N_{S}}{\sqrt{N_{S}+N_{B}+(0.25\times (N_{B}+N_{S}))}}$ as our figure of merit, where $N_{S}$ represents the expected signal yield and $N_{B}$ the total background yield. The term $0.25\times (N_{B}+N_{S})$ represents a 25\% systematic uncertainty on the signal plus background prediction, which is a realistic uncertainty based on VBF searches at ATLAS and CMS~\cite{Khachatryan:2016mbu,VBF2,Sirunyan:2019zfq, ATLASCompressedSUSY13TeV2016to2018data,CMSCompressedSUSY13TeV2016to2018data}. We note this particular definition of signal significance is only used for the purpose of optimizing the selections. The final discovery reach is determined with a shape based analysis (described later) using the full range of the transverse mass ($m_{T}(l,p_{T}^{\textrm{miss}})$) or dijet mass ($m_{jj}$) spectrum.

The VBF topology consists of two high-$p_{T}$ forward jets, in opposite sides of the detector, with a large difference in $\eta$, and a TeV scale dijet mass. This unique topology allows for the rejection of events from QCD and $V+$jets, with suppression factors of $10^{3}$--$10^{6}$ depending on the background. The large background reduction compensates the naturally smaller production cross sections for VBF processes. In addition, the requirement of two highly-energetic jets may allow the use of experimental triggers that do not constrain (or minimally constrain) the $p_{T}$ of the leptons from the chargino/neutralino decays. This is important to explore compressed mass spectra scenarios.

\begin{figure}[]
    \centering
    \includegraphics[width=0.5\textwidth]{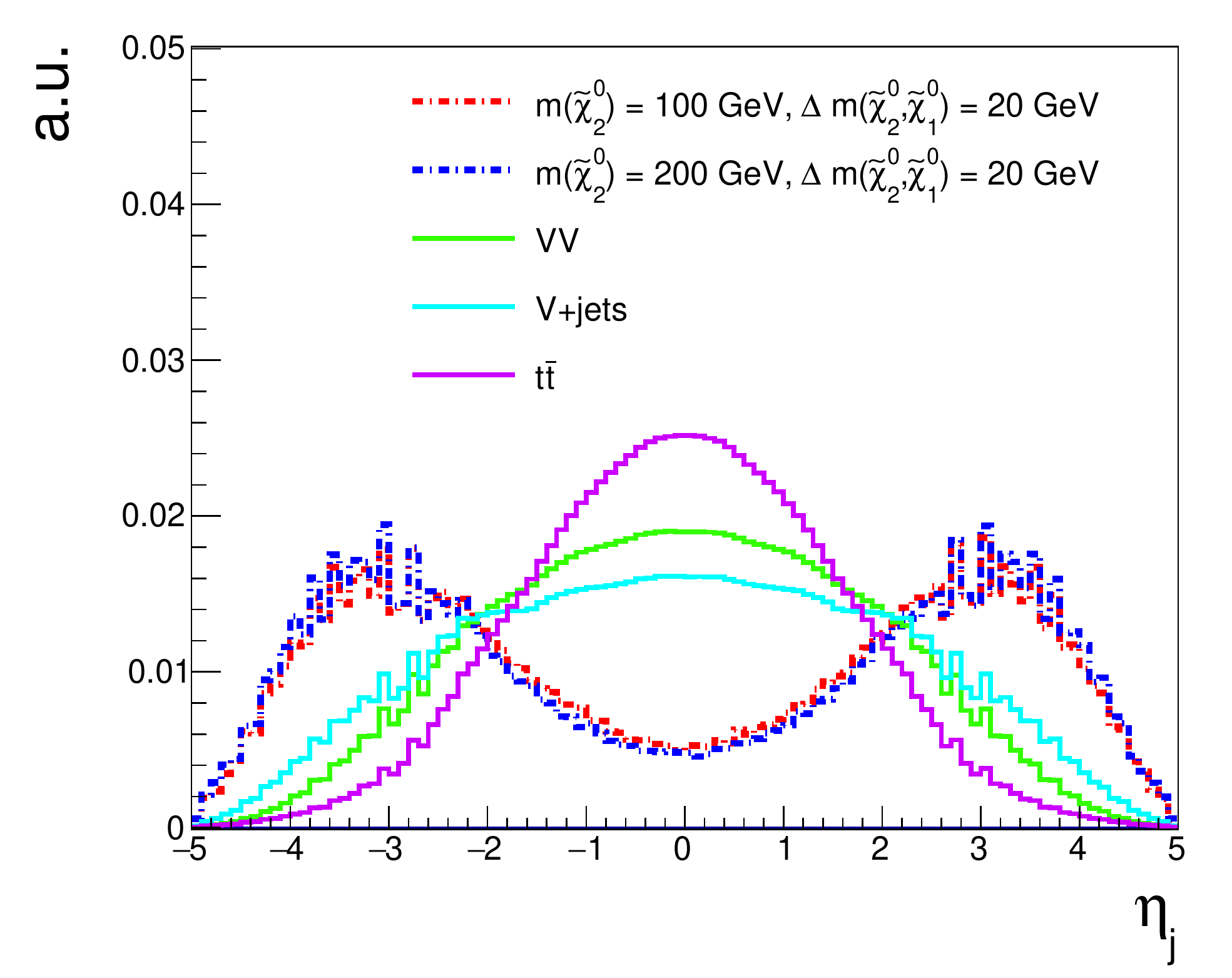}
    \caption{$\eta$ distribution for jets with $p_{T} > 30$ GeV, for two benchmark signal points and major backgrounds.}
    \label{fig:etajets}
\end{figure}

\begin{figure}[]
    \centering
    \includegraphics[width=0.49\textwidth]{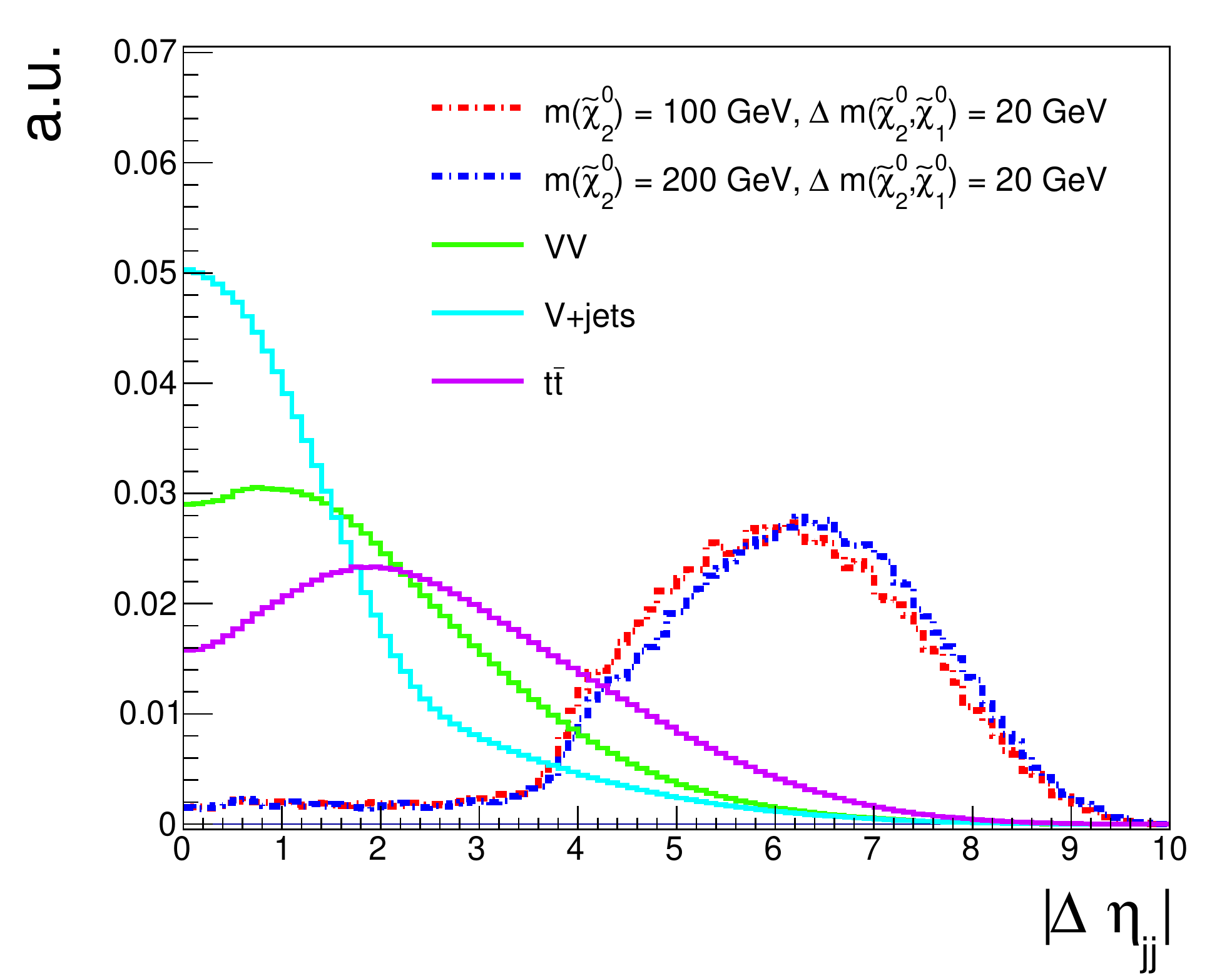}
    \caption{$|\Delta \eta_{jj}|$ distribution for dijet candidates containing jets with $p_{T} > 30$ GeV, for two benchmark signal points and major backgrounds.}
    \label{fig:deltaetajets}
\end{figure}

To illustrate important topological differences between signal and SM background processes, various kinematic distributions are presented. Since the allowed $p_{T}$ and $\eta$ phase space for reconstructed jets is limited by the experimental constraints from the ATLAS and CMS experiments, namely from effects due to the geometry of the detector, its performance, and the limitations of the jet reconstruction algorithms, distributions are studied with a pre-selection of at least two jets with $p_{T} > 30$ GeV and $|\eta| < 5$. Figures~\ref{fig:etajets} and~\ref{fig:deltaetajets} show the $\eta$ and $\Delta \eta$ distributions for jets in our major SM backgrounds and two signal benchmark samples with $\{m(\widetilde{\chi}^{0}_{2}),m(\widetilde{\chi}^{\pm}_{1}),m(\widetilde{\chi}^{0}_{1})\} = \{100 \textrm{ GeV}, 90 \textrm{ GeV}, 80 \textrm{ GeV}\}$ and  $\{200\textrm{ GeV},190\textrm{ GeV},180\textrm{ GeV}\}$. As expected, the SM backgrounds primarily contain jets with $\eta \approx 0$ that are central in the detector and that form dijet combinations with small $|\Delta\eta_{jj}|$, while the higgsino distributions are characterized by jets that travel closer to the proton beam line and form dijet pairs with large $|\Delta\eta_{jj}|$. Figure~\ref{fig:deltaetajets} motivates a stringent requirement on the pseudorapidity gap between jets, and we impose a requirement of $|\Delta\eta_{jj}| > 5.5$ as a result of the optimization process.

We note that VBF higgsino production is different from the VBF wino/bino processes studied by the ATLAS and CMS Collaborations and some of the current authors in Refs.~\cite{VBF1,DMmodels2,Khachatryan:2016mbu,VBFSlepton,VBFStop,VBFSbottom,VBF2,Sirunyan:2019zfq}. While VBF wino production occurs primarily via t-channel $WW$/$WZ$/$ZZ$ diagrams, VBF higgsino production contains important contributions from s-channel $WW$ fusion (see Figure~\ref{fig:feynVBF}). This distinguishing feature of VBF higgsino production results in more forward jets and a larger $|\Delta\eta_{jj}|$ gap in comparison to the VBF wino scenarios. This difference allows for better background suppression and also potentially the experimental differentiation of the VBF higgsino process from other SUSY scenarios. 

\begin{figure}[]
    \centering
    \includegraphics[width=0.49\textwidth]{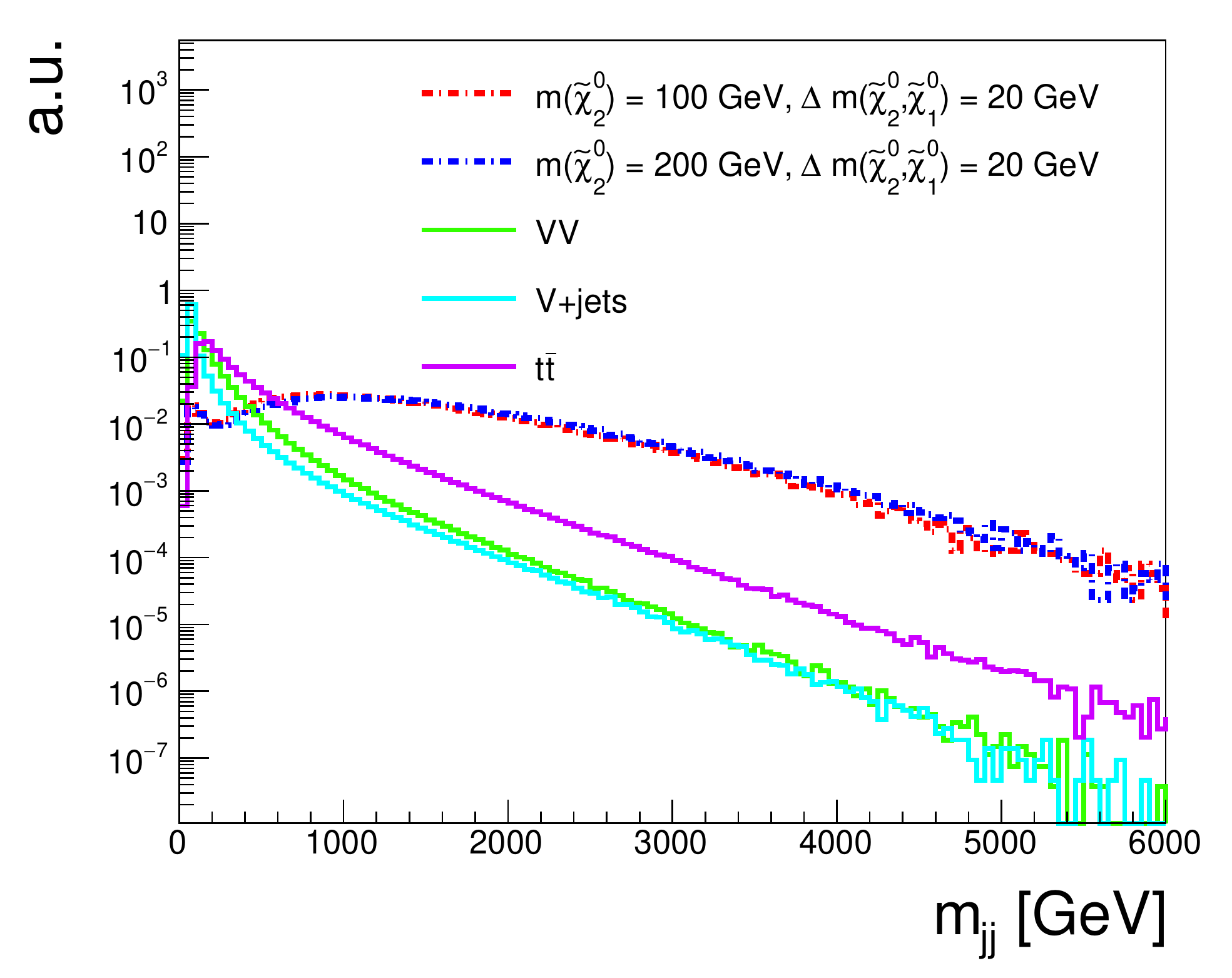}
    \caption{$m_{jj}$ distribution for jets with $p_{T} > 30$ GeV and $|\eta| < 5.0$, for two benchmark signal points and major backgrounds.}
    \label{fig:dijetmass}
\end{figure}

Figure~\ref{fig:dijetmass} shows the reconstructed $m_{jj}$ distributions (normalized to unity) of the major SM backgrounds and the signal benchmark points with $\{m(\widetilde{\chi}^{0}_{2}),m(\widetilde{\chi}^{\pm}_{1}),m(\widetilde{\chi}^{0}_{1})\} = \{100\textrm{ GeV}, 90\textrm{ GeV}, 80\textrm{ GeV}\}$ and  $\{200\textrm{ GeV},190\textrm{ GeV},180\textrm{ GeV}\}$. For events where there are more than two well reconstructed and identified jet candidates, the dijet pair with the larger value of $m_{jj}$ is used in Figure~\ref{fig:dijetmass}. At a high-energy experiment such as the LHC, where the kinetic energy of a jet is much larger than the rest energy of the parent quark, $m_{jj}$ is well-approximated by $m_{jj} \approx \sqrt{2p_{T}^{j_{1}}p_{T}^{j_{2}}\textrm{cosh}(\Delta\eta_{jj})}$. Therefore, the large $|\Delta\eta_{jj}|$ characterizing VBF higgsino production results in a broad signal distribution that overtakes the SM backgrounds at near TeV scale values. We select events with at least one dijet candidate with $m_{jj} > 0.5$ TeV. In the rare cases ($<1$\%) where selected events contain more than one dijet candidate satisfying the VBF criteria, the VBF dijet candidate with the largest value of $m_{jj}$ is chosen since it is 99\% likely to result in the correct VBF dijet pair for signal events.

Similar to current ATLAS and CMS searches for SUSY, the production of $\tilde{\chi}_{1}^{0}$ candidates at the LHC is indirectly inferred through the measurement of momentum imbalance in the transverse plane of the detectors.  The reconstructed ${p}_{T}^{\textrm{miss}}$, defined as the magnitude of the negative vector sum of the transverse momentum of visible objects, $p_{T}^{\textrm{miss}} = |-\sum_{i=\textrm{visible}}\vec{p}_{T,i}|$, is required to be greater than 175 GeV, as determined by the optimization procedure. The $p_{T}^{\textrm{miss}}$ cut is especially effective at suppressing the $V$+jets background, where the average $p_{T}^{\textrm{miss}}$ is constrained by the $Z$ or $W$ mass.  The efficiency of the $p_{T}^{\textrm{miss}}$ selection is approximately 25\%, while the SM backgrounds are reduced by approximately 2–4 orders of magnitude, depending on the process. 

\begin{figure}[]
    \includegraphics[width=0.49\textwidth]{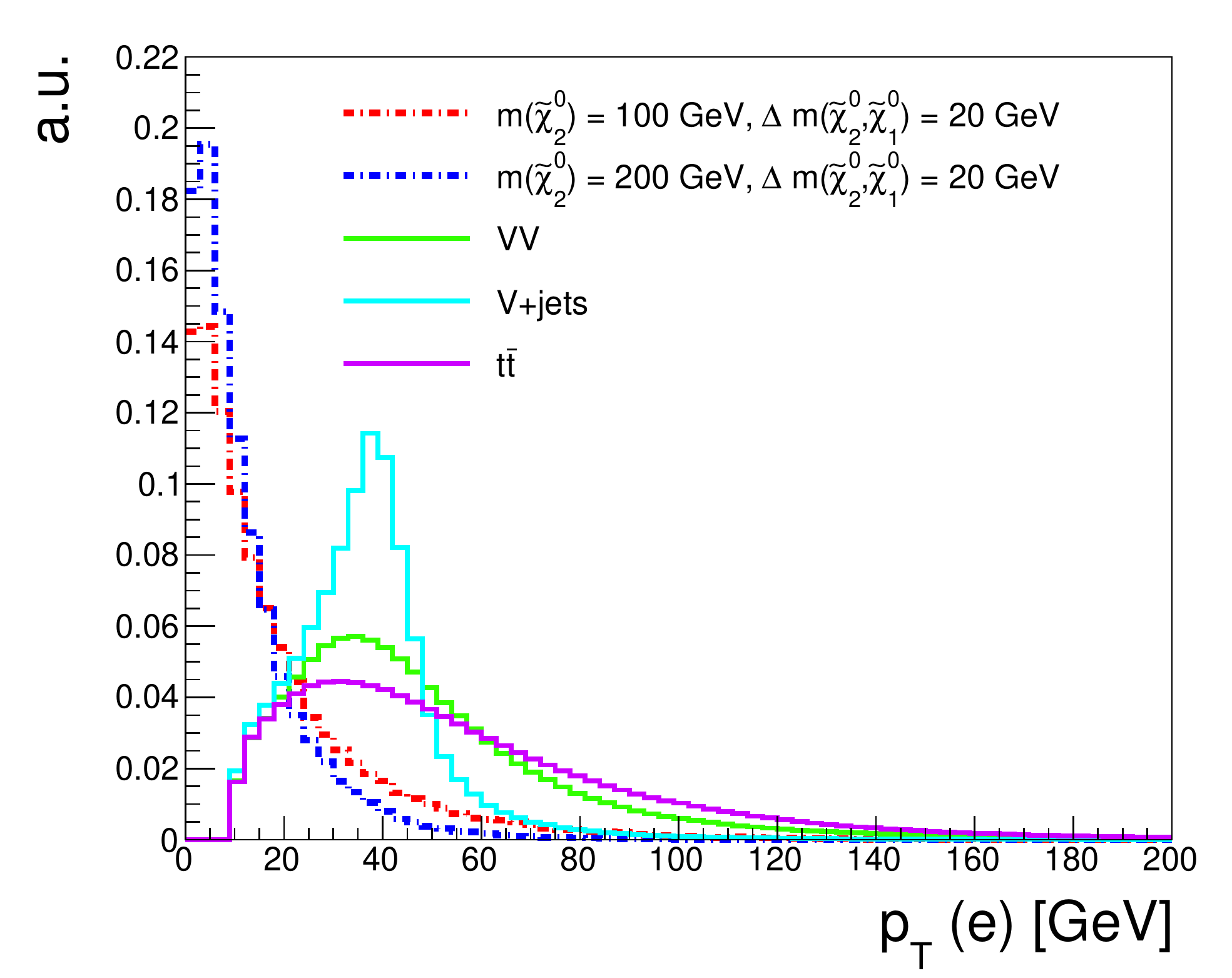}
    \caption{$p_{T}$ distribution for electrons with $p_{T} > 3$ GeV and $|\eta| < 2.5$, for two benchmark signal points and major backgrounds.}
    \label{fig:electronPt}
\end{figure}

\begin{figure}[]
    \includegraphics[width=0.49\textwidth]{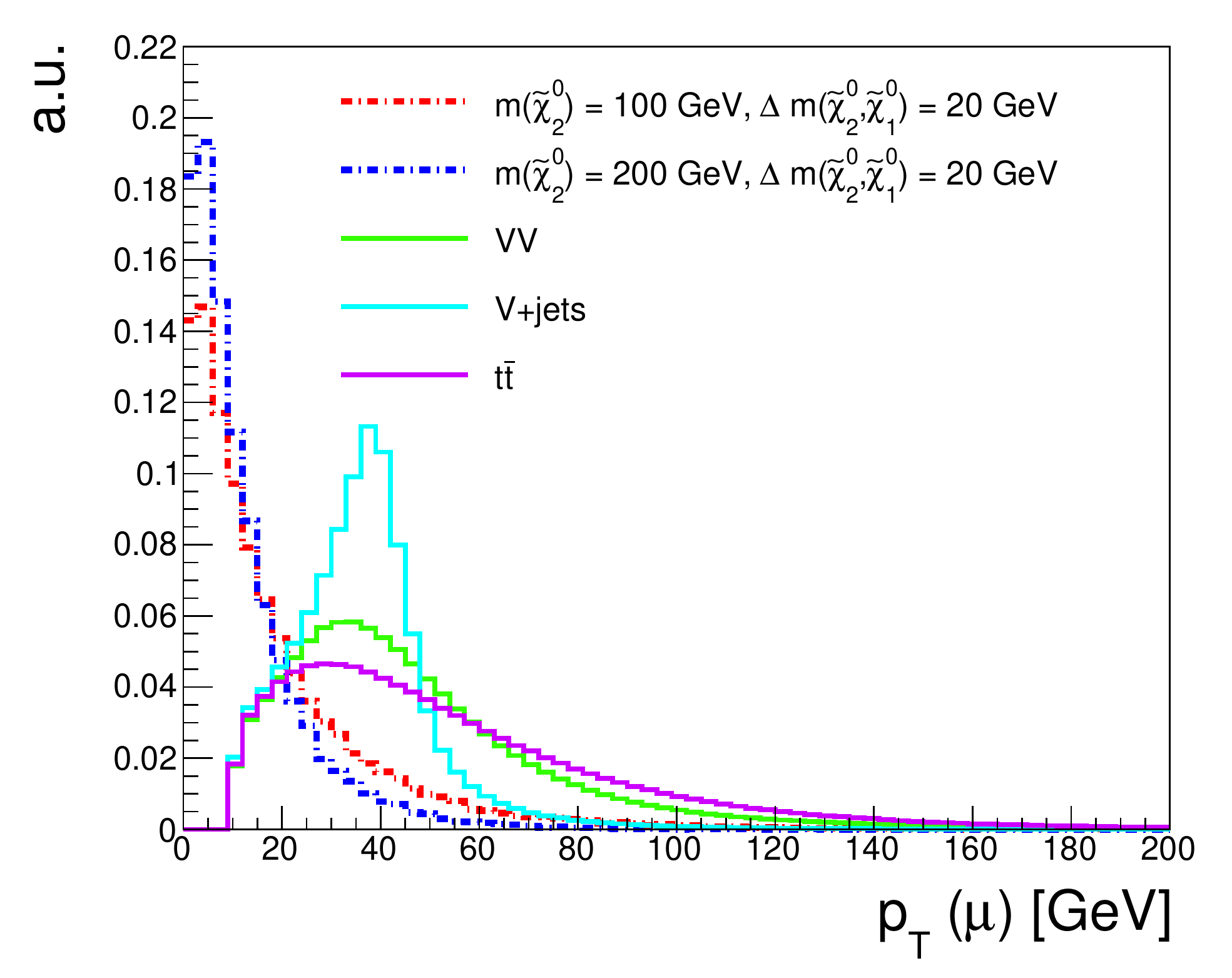}
    \caption{$p_{T}$ distribution for muons with $p_{T} > 3$ GeV and $|\eta| < 2.5$, for two benchmark signal points and major backgrounds.}
    \label{fig:muonPt}
\end{figure}

As outlined previously, besides the $p_{T}^{\textrm{miss}}$ requirement and the two oppositely directed forward jets that characterize VBF production, we require the presence of one or two light leptons. Events are classified into six search regions depending on the lepton flavor, lepton multiplicity, and the $p_{T}$ requirements on the leptons. This is motivated by the observation that the lepton multiplicity and average $p_{T}$ strongly depends on $\Delta m$. Lower (higher) values of $\Delta m$ result in a softer (harder) lepton $p_{T}$ spectrum, and consequently in a lower (larger) probability to reconstruct and identify multiple leptons. Therefore, the six search categories complement each other to provide the best discovery reach over the full phase space of $\Delta m < 50$ GeV. For example, as will be shown in Section IV, the single-lepton final state provides the best sensitivity to the $\Delta m < 20$ GeV scenarios, where the probability to reconstruct and identify multiple leptons is low. Figures~\ref{fig:electronPt} and~\ref{fig:muonPt} show the $p_{T}$ distribution for electrons and muons, respectively. Electron and muon candidates populating these distributions are pre-selected with $p_{T}(l) >$ 3 GeV and $|\eta(l)| < 2.5$. The $\eta$ requirement is driven by the geometric constraints of the CMS tracker sub-detector, while the lower thresholds on electron/muon $p_{T}$ are motivated by a combination of experimental constraints on lepton reconstruction/identification and achieving good signal significance. As shown in Figures~\ref{fig:electronPt} and~\ref{fig:muonPt}, the leptons from signal processes have an average transverse momentum of $p_{T}(l) \sim \Delta m / 3$, while the SM backgrounds have an average lepton $p_{T}$ of $m_{W}/2$ or $m_{Z}/2$. We take advantage of this characteristic by imposing an upper threshold of $p_{T}(l) < m_{W}/2$. The single electron channel contains two search categories, where electrons are required to have $8 < p_{T}(e) < 40$ GeV or $8 < p_{T}(e) < 15$ GeV, targeting optimal sensitivity for different $\Delta m$ values. The lower $p_{T}$ threshold is determined through the optimization procedure (e.g. the jet$\to e$ misidentification rate is high for $p_{T}(e) \sim 3$ GeV). Similarly, the single muon channel contains two search categories with thresholds set to $5 < p_{T}(\mu) < 40$ or $5 < p_{T}(\mu) < 15$. For the two dilepton channels, an upper threshold of $p_{T}(l) < 40$ GeV is applied. 

To further suppress SM backgrounds with top quarks, we impose a b-jet veto requirement. Events are rejected if a jet with $p_{T} > 30$ GeV and $|\eta|< 2.4$ is identified as a bottom quark ($b$). Events are also rejected if they contain jets with $p_{T} > 20$ GeV and $|\eta|< 2.5$ tagged as hadronically decaying tau leptons. The $\tau_{h}$ veto requirement further reduces backgrounds with vector boson pairs, while being $> 95$\% efficient for VBF higgsino signal events. Table~\ref{tab:selection_criteria} summarizes the proposed event selection criteria. Note that the selections have been separated into different sets identified with Roman numbers. For example, the jet pre-selections are identified as set I, while the VBF selections are identified as set III. These Roman numbers will be used in the tables that follow.

\hspace{-1cm}

\begin{table}[]
\begin{center}
\caption {Event selection criteria to maximize discovery potential at the LHC, for the single-lepton (single electron and single muon) and the dilepton (dielectron and dimuon) channels.}
\label{tab:selection_criteria}
\begin{tabular}{ l  c r}\hline\hline
Criterion & $e/\mu jj$\\
 \hline
  \multicolumn{3}{ c }{{\bf Initial jet selections (I)}} \\
  \hline
   $p_{T}(j)$ & & $>30$ GeV\\
   $|\eta(j)|$ & & $< 5.0$\\
   $\Delta R(j_{1},j_{2})$ & & $> 0.3$\\ 
  \hline
  \multicolumn{3}{ c }{{\bf Topological selections (II)}} \\
  \hline
  
  $|\eta($b-jets$)|$ &  & $< 2.5$\\
  $p_{T}($b-jets$)$ & & $> 30$ GeV\\
  $N($b-jets$)$ & & $= 0$\\
  $|\eta(\tau_{h})|$ &  & $< 2.5$\\
  $p_{T}(\tau_{h})$ & & $> 20$ GeV\\
  $N(\tau_{h})$ & & $= 0$\\
  $p_{T}^{\textrm{miss}}$ & & $> 150$ GeV \\
   \hline
   \multicolumn{3}{ c }{{\bf VBF selections (III)}} \\
   \hline
   
   $N(j)$ & & $\geq$ 2\\
   $\eta(j_{1})\cdot \eta(j_{2})$ & & $< 0$\\
   $|\Delta \eta (j_{1}, j_{2})| $ & & $> 5.5$\\
   
   \hline
   \multicolumn{3}{ c }{{\bf Single electron/muon (IV/V)}} \\
   \hline
   
   $N(e/\mu)$ &   &  $= 1$ \\
  $|\eta(e/\mu)|$ &  & $< 2.4$\\ 
  $p_{T}(e/\mu)$ & & $> 8/5$ GeV \& $< 15$ GeV\\
     \hline
   \multicolumn{3}{ c }{{\bf Single electron/muon (VI/VII)}} \\
   \hline
   
   $N(e/\mu)$ &   &  $= 1$ \\
  $|\eta(e/\mu)|$ &  & $< 2.4$\\ 
  $p_{T}(e/\mu)$ & & $> 8/5$ GeV \& $< 40$ GeV\\
   
   \hline
   \multicolumn{3}{ c }{{\bf Dielectron/Dimuon (VIII/IX)}} \\
   \hline
   
   $N(e/\mu)$ &   &  $= 2$ \\
  $|\eta(e/\mu)|$ &  & $< 2.4$\\ 
  $p_{T}(e/\mu)$ & & $> 8/5$ GeV \& $< 40$ GeV\\
   \hline\hline
 \end{tabular}
\end{center}
\end{table}

\begin{figure*}[!htb]
    \centering
        \subfloat{{\includegraphics[width=8.5cm]{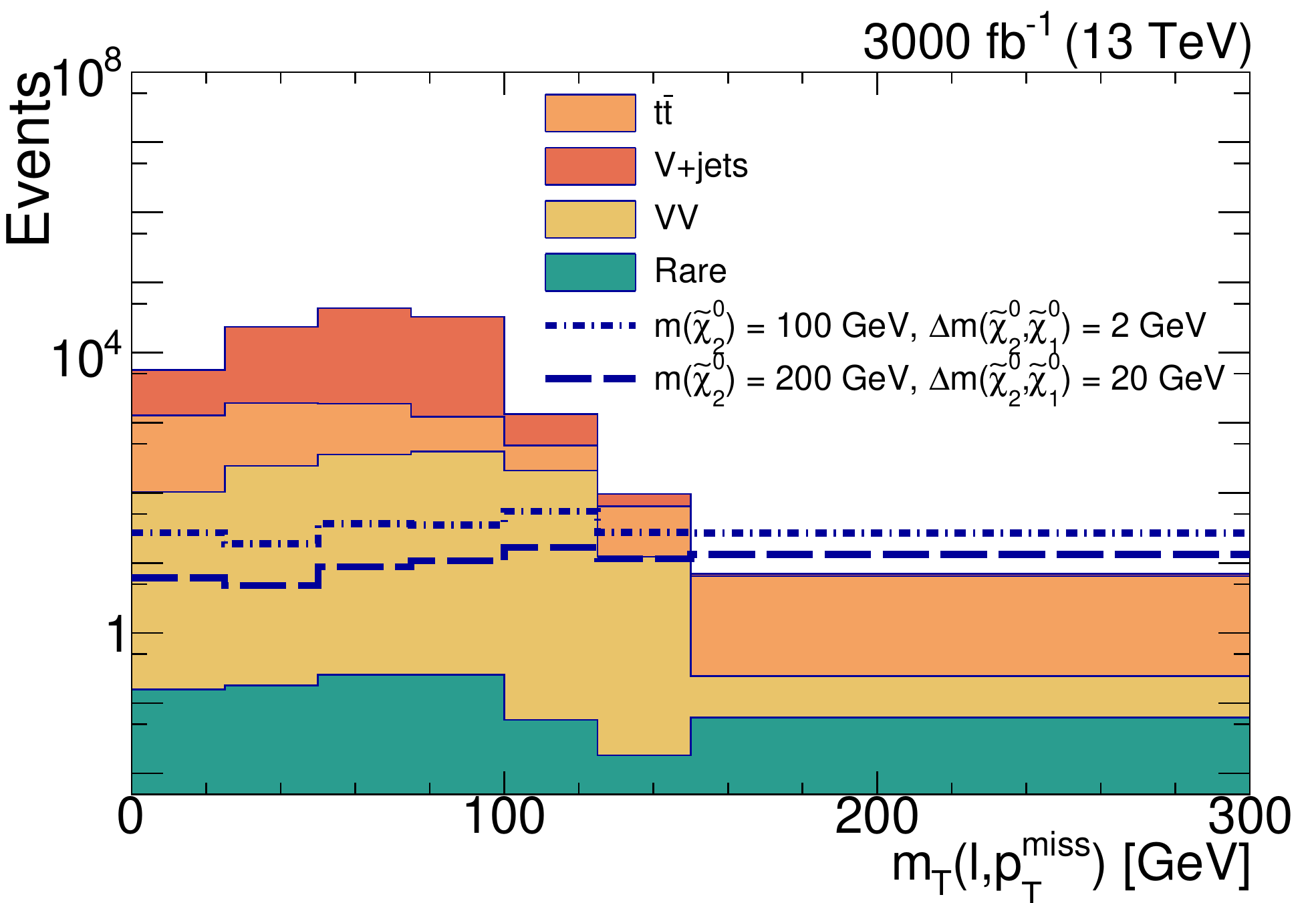}}}%
    \qquad
    \subfloat{{\includegraphics[width=8.5cm]{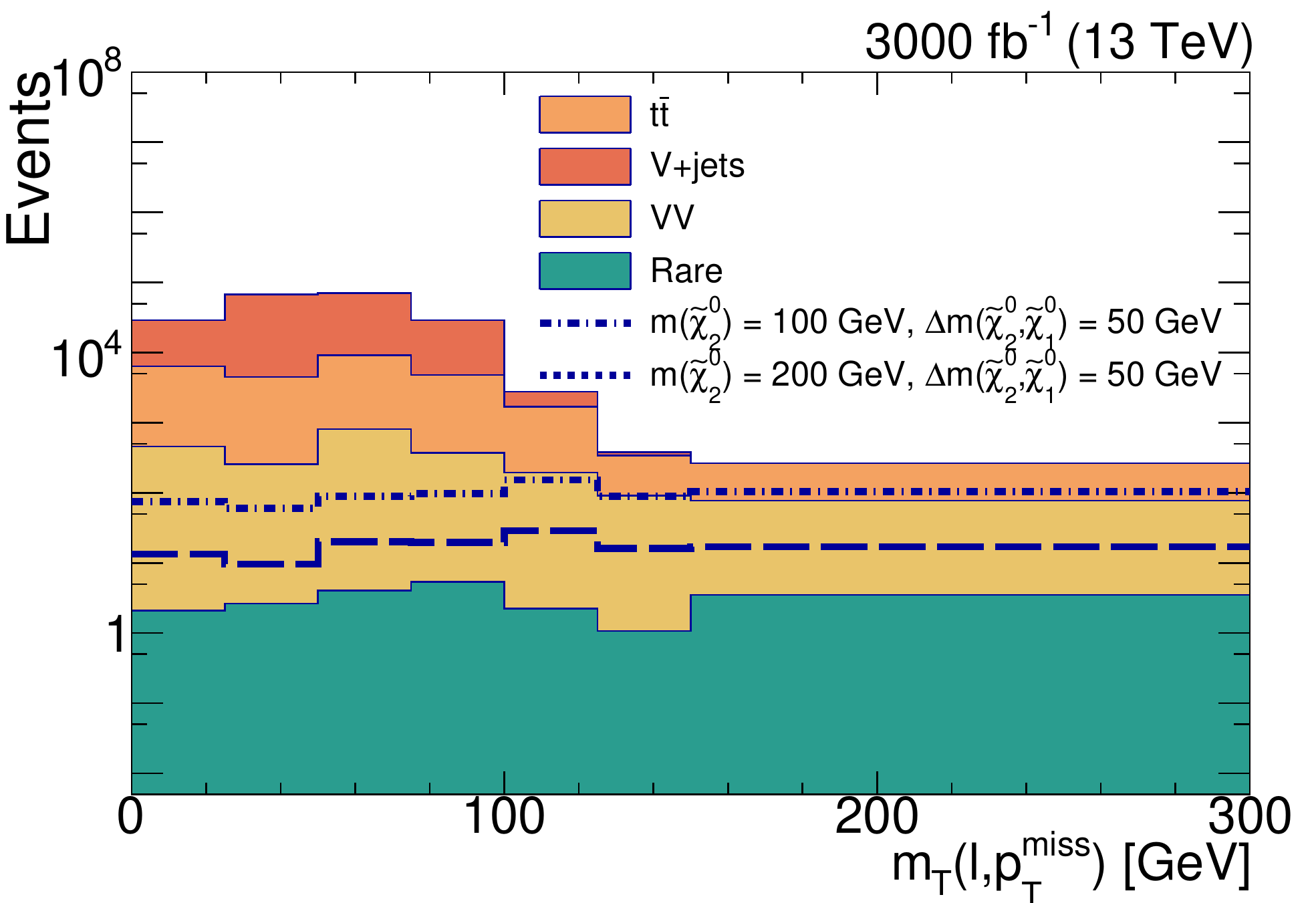} }}%
    \caption{The $m_{T}$ distributions in the single-lepton signal regions with $p_{T}(l) < 15$ GeV (left) and $p_{T}(l) < 40$ GeV (right).}%
    \label{fig:StackedmT}%
\end{figure*}

\begin{figure}[]
    \includegraphics[width=0.49\textwidth]{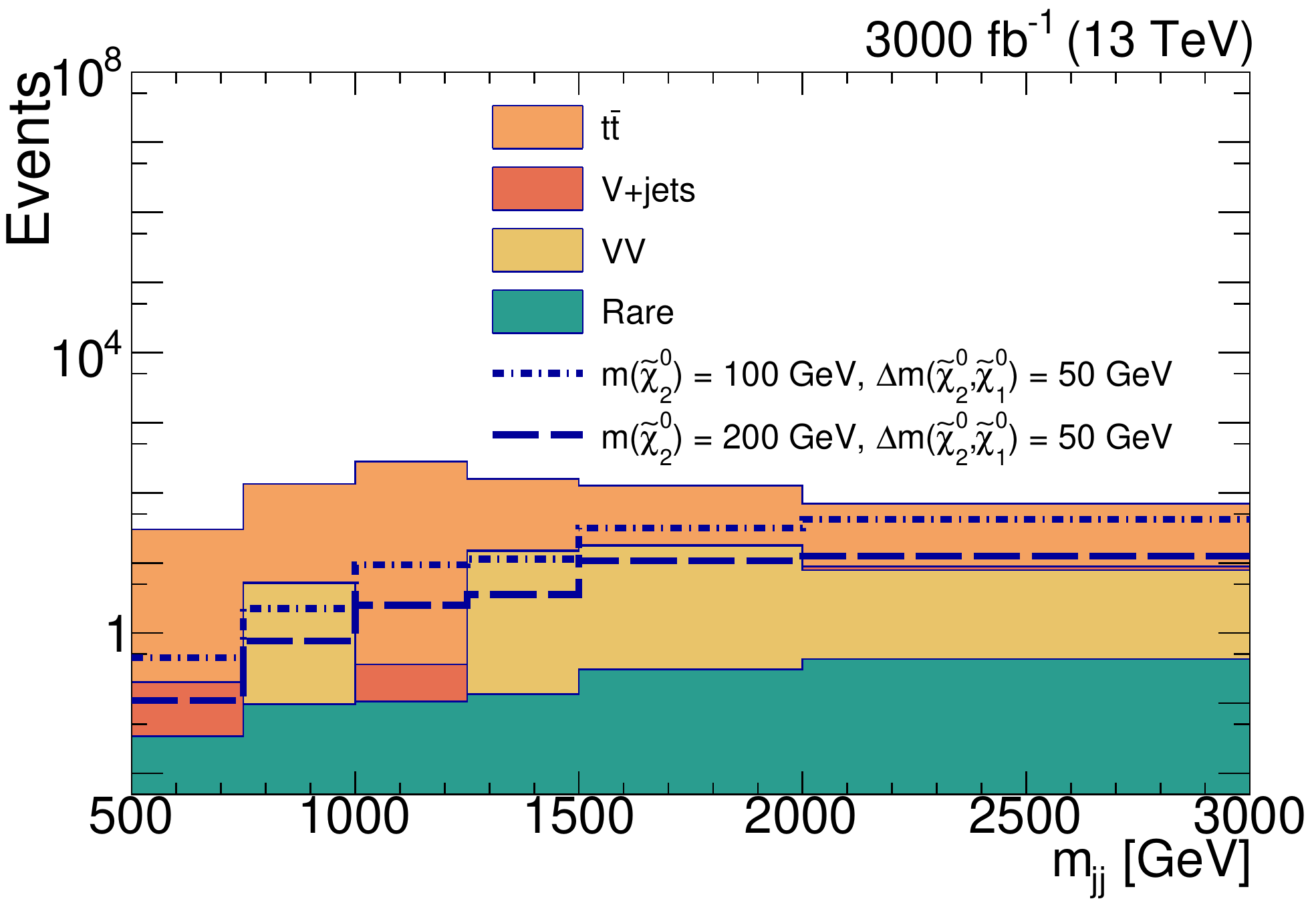}
    \caption{The $m_{jj}$ distribution in the di-lepton signal region.}
    \label{fig:Stackedmjj}
\end{figure}

Table~\ref{tab:efficiencies} shows the cumulative efficiencies 
for the background samples and some benchmark signal samples, presented as percentages. Note that the cut efficiencies for the SM backgrounds are several orders of magnitude smaller in comparison to the signal acceptances. Figure~\ref{fig:StackedmT} shows the $m_{T}$ distribution, after all the event selection criteria, for the single-lepton signal regions with $p_{T}(l) < 15$ GeV (left) and $p_{T}(l) < 40$ GeV (right). Figure~\ref{fig:StackedmT} (left) corresponds to selections IV$+$V in Table~\ref{tab:selection_criteria}, while Fig.~\ref{fig:StackedmT} (right) corresponds to selections VI$+$VII. Figure~\ref{fig:Stackedmjj} shows the expected background and signal yields in bins of $m_{jj}$ for the dilepton signal region (selections VIII$+$IX in Table~\ref{tab:selection_criteria}). In Figs.~\ref{fig:StackedmT}--\ref{fig:Stackedmjj}, the signal distributions are overlaid on the background distributions, which are stacked on top of each other. Figure~\ref{fig:Stackedmjj} shows the reconstructed dijet mass ($m_{jj}$) distribution for the di-lepton channel. The expected number of events for the signal benchmark samples and SM backgrounds in Figs.~\ref{fig:StackedmT}--\ref{fig:Stackedmjj} are normalized to cross section times the assumed integrated luminosity of 3000 fb$^{-1}$

\begin{table*}[!htb]
\centering
\caption{Cumulative efficiencies for some signal and backgrounds samples, expressed as percentages. The Roman numbers are associated to the segments of events selections presented in Table~\ref{tab:selection_criteria}.}
\label{tab:efficiencies}%
\setlength{\tabcolsep}{3pt}
\begin{tabular}{c | c | c | c | c | c | c | c}
\hline \hline 
& \multicolumn{1}{c|}{Sample}& \multicolumn{1}{c|}{I}& \multicolumn{1}{c|}{II}& \multicolumn{1}{c|}{III}& \multicolumn{1}{c|}{IV+V}& \multicolumn{1}{c|}{VI+VII}& \multicolumn{1}{c}{VIII+IX}\\ \hline
 \multirow{5}{*}{\STAB{\rotatebox[origin=c]{90}{signals}}} 
& $\{m(\widetilde{\chi}^{0}_{2}),m(\widetilde{\chi}^{\pm}_{1}),m(\widetilde{\chi}^{0}_{1})\} = \{100, 75, 50\}$   & 61.2    &22.5   &12.3      &0.20       &1.08      &0.20\\ 
& $\{m(\widetilde{\chi}^{0}_{2}),m(\widetilde{\chi}^{\pm}_{1}),m(\widetilde{\chi}^{0}_{1})\} = \{100, 85, 70\}$   & 60.6    &19.1   &9.9       &0.36       &1.60      &0.37\\ 
& $\{m(\widetilde{\chi}^{0}_{2}),m(\widetilde{\chi}^{\pm}_{1}),m(\widetilde{\chi}^{0}_{1})\} = \{200, 175, 150\}$ & 61.3    &20.8   &11.1      &0.38       &1.65      &0.52\\ 
& $\{m(\widetilde{\chi}^{0}_{2}),m(\widetilde{\chi}^{\pm}_{1}),m(\widetilde{\chi}^{0}_{1})\} = \{200, 185, 170\}$ & 60.3    &18.0   &9.3       &0.51       &2.09      &0.62\\ 
& $\{m(\widetilde{\chi}^{0}_{2}),m(\widetilde{\chi}^{\pm}_{1}),m(\widetilde{\chi}^{0}_{1})\} = \{300, 285, 270\}$ & 59.4    &17.4   &9.2       &0.69       &2.38      &0.81\\ 
\hline 
 \multirow{6}{*}{\STAB{\rotatebox[origin=c]{90}{backgrounds}}} 
&$Z$+jets         &33.9        &0.06     &1.49    $\times 10^{-3}$     &5.40     $\times 10^{-6}$        &5.40  $\times 10^{-6}$      &$7.73$ $\times 10^{-10}$\\ 
&$W$+jets         &27.1        &0.03     &9.60    $\times 10^{-4}$     &1.91    $\times 10^{-5}$        &4.30  $\times 10^{-5}$      &2.25  $\times 10^{-10}$\\ 
&$t\bar{t}$+jets     &81.1        &0.53     &2.37    $\times 10^{-2}$     &3.04    $\times 10^{-4}$        &1.64  $\times 10^{-3}$      &4.11    $\times 10^{-5}$\\ 
&$WW$+jets             &54.4        &0.56     &1.08    $\times 10^{-2}$     &1.76    $\times 10^{-4}$        &6.56  $\times 10^{-4}$      &1.60    $\times 10^{-5}$\\ 
&$ZZ$+jets             &59.4        &0.98     &1.49    $\times 10^{-2}$     &2.00    $\times 10^{-5}$        &1.20  $\times 10^{-4}$      &2.00    $\times 10^{-5}$\\ 
&$WZ$+jets             &57.2        &0.87     &1.88    $\times 10^{-2}$     &1.42    $\times 10^{-4}$        &5.18  $\times 10^{-4}$      &8.22    $\times 10^{-6}$\\ 
\hline \hline 
\end{tabular}
\end{table*}

\section{Results}\label{sec:Results}

As noted, the signal significance 
described in the previous section is only used to optimize the selections. However, instead of a cut and count approach, maximum likelihood fits are performed using the full range of the $m_{T}$ and $m_{jj}$ distributions to construct a profile likelihood ratio test statistic, and subsequently determine the discovery reach. The test statistic was constructed using the ROOT-Fit \cite{ROOTFit} toolkit package developed by CERN. The expected bin-by-bin yields of the $m_{T}$ and $m_{jj}$ distributions in Figures~\ref{fig:StackedmT}--\ref{fig:Stackedmjj}, obtained using events satisfying the selections in Table~\ref{tab:selection_criteria}, are used as input to the profile binned likelihood calculation. Systematic uncertainties are incorporated in the significance calculation via nuisance parameters, assuming log-normal and Gaussian priors for normalization and shape uncertainties, respectively. The value of the significance is determined using the measured local p-value, calculated as the probability under a background only hypothesis to obtain a value of the test statistic as large as that obtained with a signal plus background hypothesis. Then, the shape based signal significance $z_{\textrm{fit}}$ is obtained by calculating the value at which the integral of a Gaussian between $z_{\textrm{fit}}$ and $\infty$ matches the local p-value.

The following systematic uncertainties were included in the calculation of $z_{\textrm{fit}}$. A 12\% uncertainty was assigned to account for differences in simulation associated with the chosen set of parton distribution functions (PDFs), used to produce the signal and background samples. These uncertainties were calculated following the PDF4LHC prescription~\cite{Butterworth:2015oua}. The effect of the chosen PDF set on the shape of the $m_{T}$ and $m_{jj}$ distributions is negligible. The uncertainty related with reconstruction and identification of forward (high-$\eta$) jets, which has a direct impact on the VBF cut efficiency and subsequently on the expected number of signal and background events, was assigned a total value of 20\%, based upon Refs.~\cite{Sirunyan:2018ubx,Aaboud:2018jiw,Aaboud:2018sua,Sirunyan:2018iwl}. Finally, a conservative 10\% uncertainty was included to account for the efficiencies of soft electron/muon reconstruction and identification requirements. The uncertainties between the two leptons in the dilepton channels, and between signal and background processes, were considered to be fully correlated. Finally, we consider shape based uncertainties due to how well jet energies are reconstructed in the forward regions of the detector. These uncertainties are additionally propagated as effects on the reconstruction of $p_{T}^{\textrm{miss}}$. The jet energy uncertainties directly affect the uncertainties on $m_{jj}$, while they indirectly affect the uncertainties on $m_{T}$ (via $p_{T}^{\textrm{miss}}$). The shape based uncertainties range from 2-10\%, depending on the $m_{T}$ or $m_{jj}$ bin.

\begin{figure}[]
    \centering
    \includegraphics[width=0.45\textwidth]{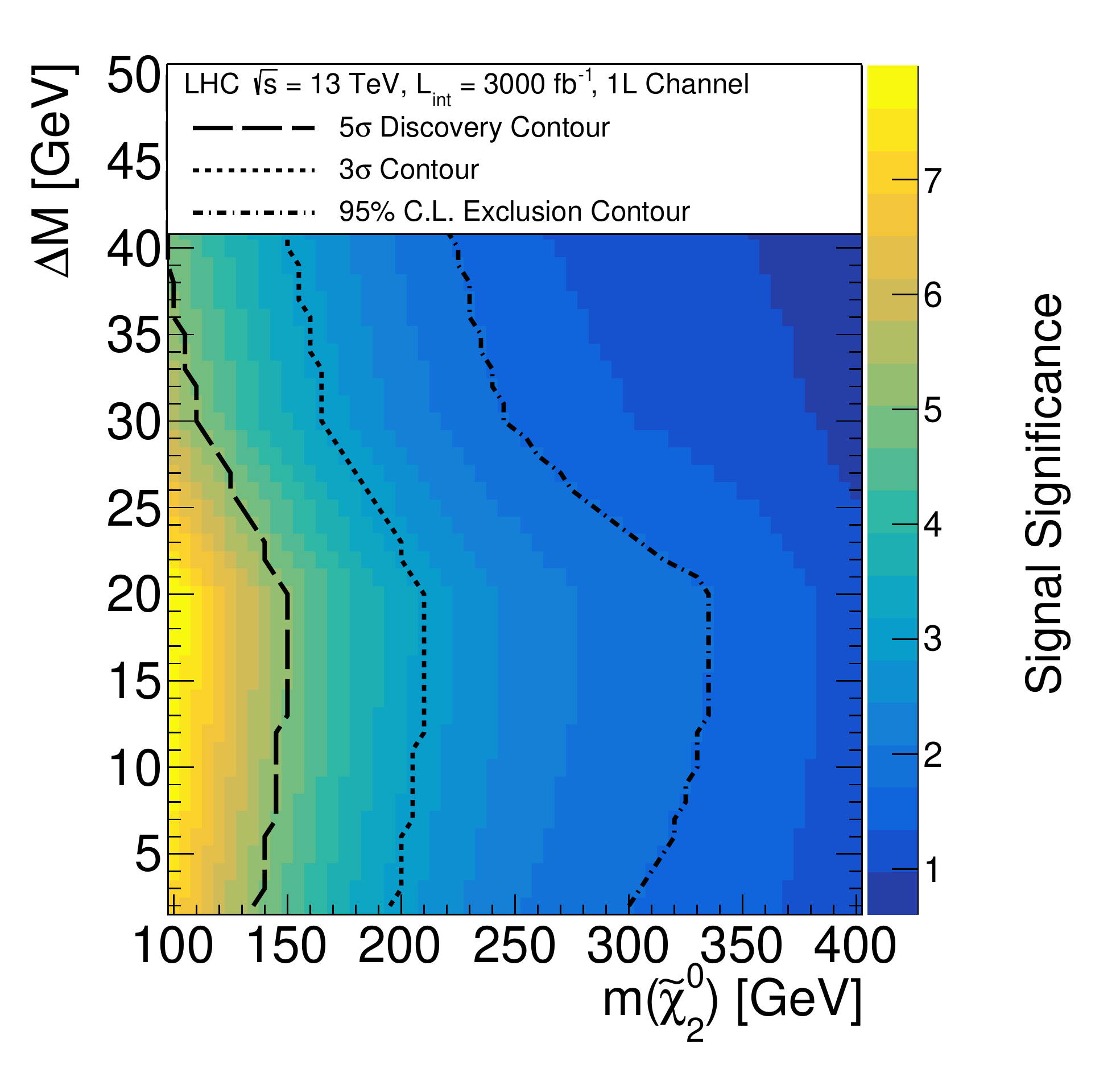}
    \caption{Combined signal significance for the single lepton final states as function of $\tilde{\chi}^{0}_{2}$ mass, for different integrated luminosity scenarios.}
    \label{fig:s_singlelepton}
\end{figure}

\begin{figure}[]
    \centering
    \includegraphics[width=0.45\textwidth]{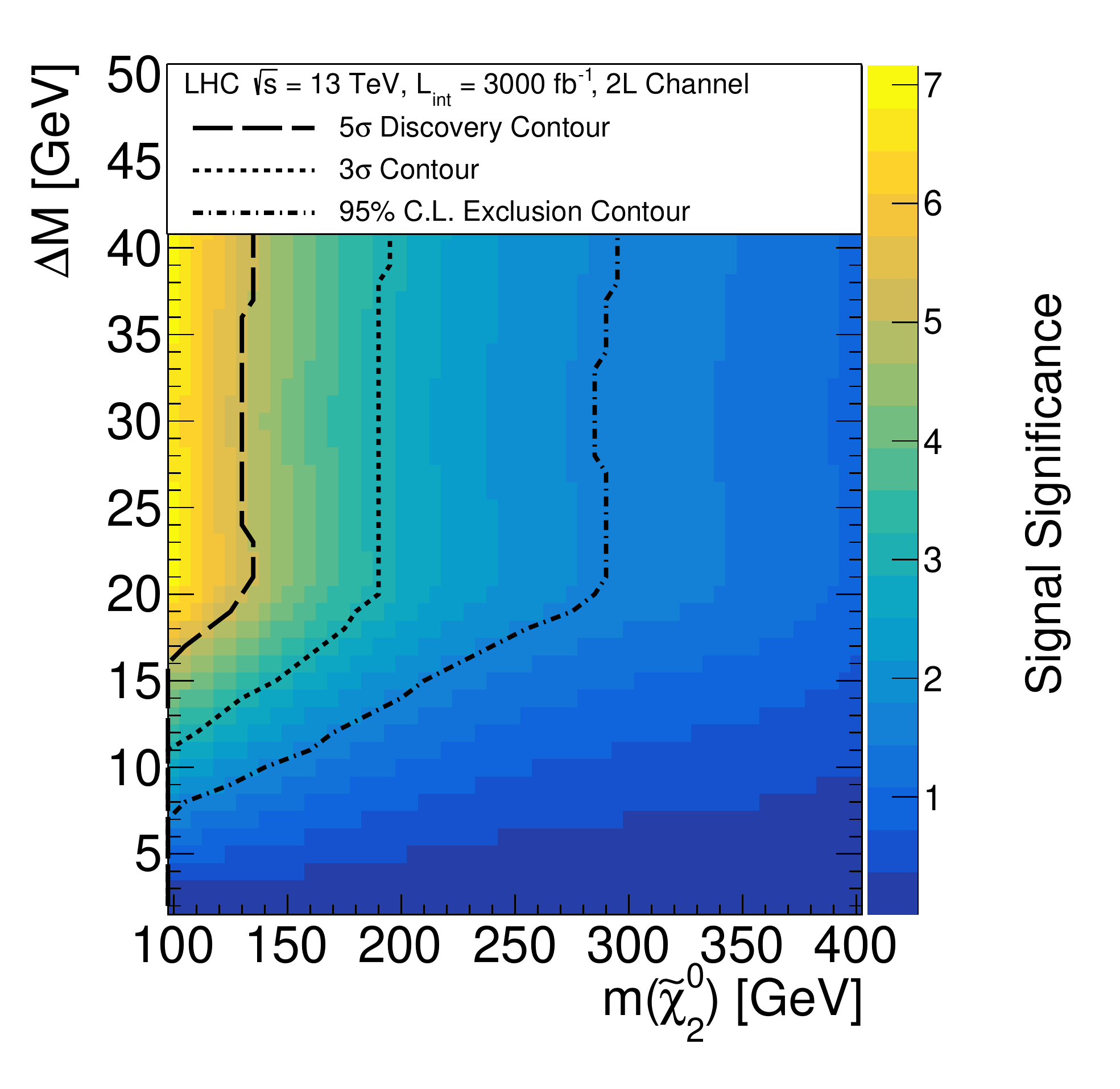}
    \caption{Combined signal significance for the dilepton final states as function of $\tilde{\chi}^{0}_{2}$ mass, for different integrated luminosity scenarios.}
    \label{fig:s_dileptont}
\end{figure}

\begin{figure}[]
    \centering
    \includegraphics[width=0.45\textwidth]{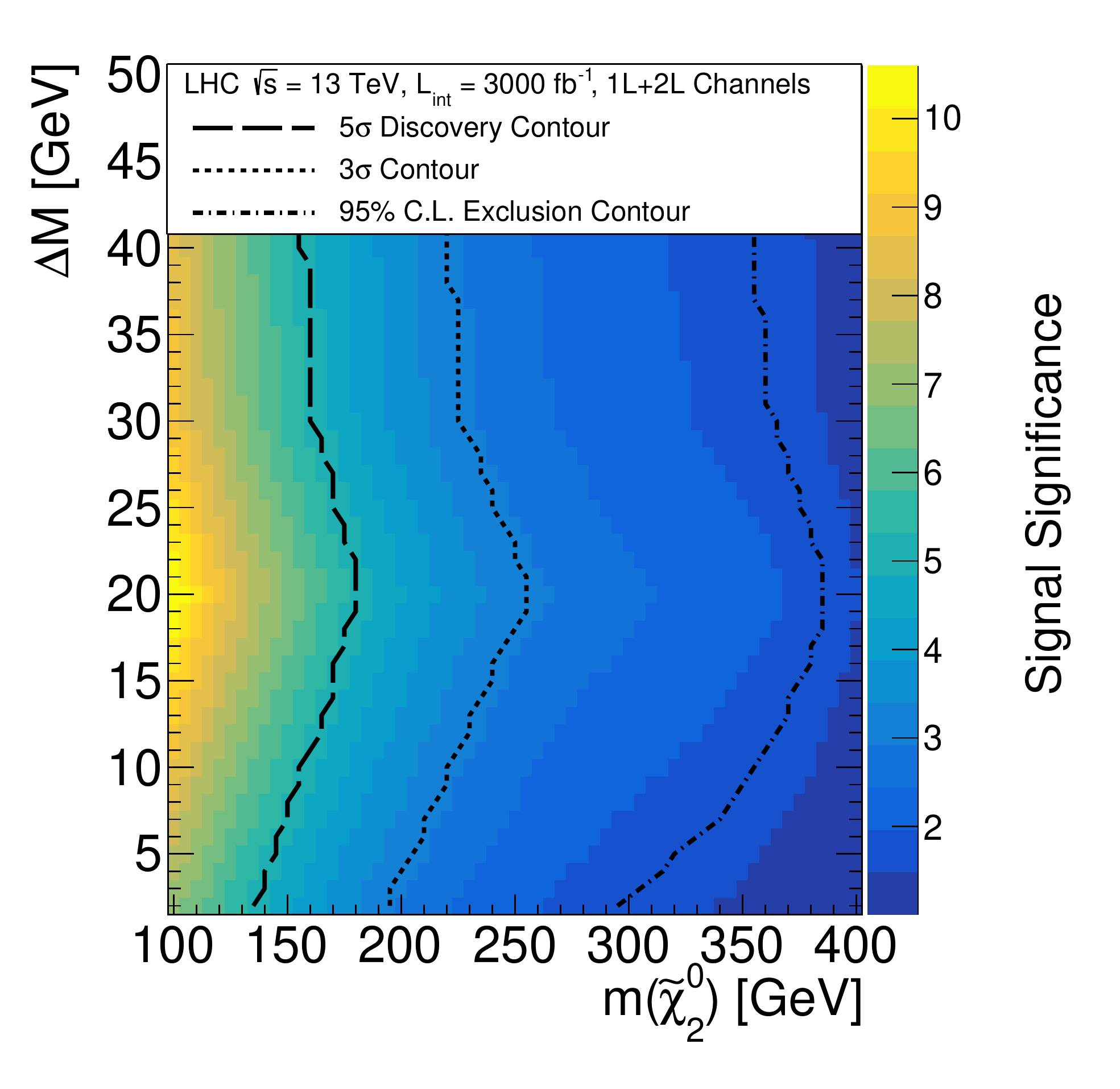}
    \caption{Total combined signal significance, including the single lepton and dilepton final states as function of $\tilde{\chi}^{0}_{2}$ mass, for different integrated luminosity scenarios.}
    \label{fig:s_combined}
\end{figure}

Figure~\ref{fig:s_singlelepton}, shows the expected signal significance (on the z-axis), as function of $m(\tilde{\chi}^{0}_{2})$ and $\Delta m(\tilde{\chi}^{0}_{2},\tilde{\chi}^{0}_{1})$ on the $xy$-plane. We assume an integrated luminosity of 3000 fb$^{-1}$ expected by the end of the HLHC era. The dashed lines delimit the 5$\sigma$ discovery region, $3\sigma$ contour, and the projected 95\% confidence level (CL) exclusion contour (should there be no evidence of an excess). Considering only the single lepton channels, there is 5$\sigma$ discovery potential for $m(\tilde{\chi}^{0}_{2})$ up to 150 GeV, assuming $\Delta m(\tilde{\chi}^{0}_{2},\tilde{\chi}^{0}_{1}) < 20$ GeV. As noted previously, Fig.~\ref{fig:s_singlelepton} shows that the VBF single lepton channels are effective probes for small $\Delta m$ values, but do not show discovery potential beyond the LEP bounds for $\Delta m$ values approaching 40-50 GeV. This is because the average lepton $p_{T}$ becomes larger at those $\Delta m$ values, and thus it becomes preferable to select events with two leptons. Figure~\ref{fig:s_dileptont} shows the expected signal significance for the combination of the dilepton channels. Similar to Fig.~\ref{fig:s_singlelepton}, the dashed lines delimit the discovery contours, as well as the projected exclusion region. The dilepton channels are effective probes of the larger $\Delta m$ phase space, resulting in a 5$\sigma$ (3$\sigma$) reach for $m(\tilde{\chi}^{0}_{2})$ up to 140 (200) GeV. Figure~\ref{fig:s_combined} presents the expected signal significance, as a function of $m(\tilde{\chi}^{0}_{2})$ and $\Delta m$, for the combination of all the single lepton and dilepton channels. Combining the search channels allows us to have sensitivity to a broader range of phase space, providing 5$\sigma$ (3$\sigma$) reach for $m(\tilde{\chi}^{0}_{2}) < 180$ $(260)$ GeV, and a projected 95\% CL exclusion region that covers $m(\tilde{\chi}^{0}_{2})$ up to 385 GeV. The expected discovery and exclusion reach using the unique VBF topology with one or two soft leptons includes sensitivity to a regime of the compressed higgsino-like parameter space that is unconstrained by any other experiment.

\section{Discussion}

In this work, we have presented a feasibility study to assess the long-term discovery potential for higgsino-like states in compressed Supersymmetry models, considering VBF processes at the 13 TeV CERN LHC. We find that s-channel $WW$ fusion with a Higgs mediator, provides an important contribution to the overall VBF higgsino-like production cross section. This characteristic leads to a VBF higgsino topology with more forward jets and a larger dijet pseudorapidity gap compared to the VBF ``wino/bino'' process considered by the ATLAS and CMS Collaborations in Refs.~\cite{Sirunyan:2018ubx,Aaboud:2018jiw,Aaboud:2018sua,Sirunyan:2018iwl}, which is primarily dominated by t-channel $WW$/$ZZ$/$WZ$ fusion diagrams. This distinguishing feature provides a nice handle to facilitate large background suppression in order to compensate for the relatively small VBF higgsino cross sections. We have shown that these stringent VBF requirements, combined with large missing momentum and one or two low-$p_{T}$ electrons/muons, is effective at reducing the major SM backgrounds, leading to a 5$\sigma$ (3$\sigma$) discovery reach for $m(\tilde{\chi}^{0}_{2}) < 180$ $(260)$ GeV, and a projected 95\% confidence level exclusion region that covers $m(\tilde{\chi}^{0}_{2})$ up to 385 GeV, assuming an integrated luminosity of 3000 fb$^{-1}$. For comparison, prior higgsino searches from the CMS and ATLAS Collaborations have not exceeded the constraints established by the LEP experiments for $\Delta m(\tilde{\chi}^{0}_{2},\tilde{\chi}^{0}_{1}) < 3$ GeV, while the lower limit on $m(\tilde{\chi}^{0}_{2})$ is at 193 GeV for $\Delta m(\tilde{\chi}^{0}_{2},\tilde{\chi}^{0}_{1}) = 9.3$ GeV~\cite{ATLASCompressedSUSY13TeV2016to2018data,CMSCompressedSUSY13TeV2016to2018data}. Therefore, the proposed methodology using a stringent VBF topology with soft leptons can probe regions of parameter space that are currently unconstrained by other searches.

\section{Acknowledgements}

We thank the constant and enduring financial support received for this project from the faculty of science at Universidad de los Andes (Bogot\'a, Colombia), and by the Foundation for the
promotion of research and technology of Bank of the Republic
of Colombia under project number 4262, the Physics \& Astronomy department at Vanderbilt University and the US National Science Foundation. This work is supported in part by NSF Award PHY-1806612.

\end{document}